# Highly-sensitive fire alarm system based on cellulose paper with low-temperature response and wireless signal conversion

Xiaolu Li[a,b], José Sánchez del Río Saez[a,c], Xiang Ao[a,b], Abdulmalik Yusuf[a,b], De-Yi Wang[a,*]

[a.] *IMDEA Materials Institute, C/Eric Kandel, 2, 28906 Getafe, Madrid, Spain*

[b.] *E.T.S. de Ingenieros de Caminos, Universidad Politécnica de Madrid, Calle Profesor Aranguren 3, 28040 Madrid, Spain*

[c.] *E.T.S. de Ingenieria y deiseno Industrial, Universidad Politécnica de Madrid, Ronda de Valencia 3, 28012 Madrid, Spain*

Email: deyi.wang@imdea.org

**Abstract**

Highly sensitive smart sensors for early fire detection with remote warning capabilities are urgently required to improve the fire safety of combustible materials in diverse applications. The highly-sensitive fire alarm can detect fire situation within a short time quickly when a fire disaster is about to occur, which is conducive to achieve fire tuned. Herein, a novel fire alarm is designed by using flame-retardant cellulose paper loaded with graphene oxide (GO) and two-dimensional titanium carbide ($Ti_3C_2$, MXene). Owing to the excellent temperature dependent electrical resistance switching effect of GO, it acts as an electrical insulator at room temperature and becomes electrically conductive at high temperature. During a fire incident, the partial oxygen-containing groups on GO will undergo complete removal, which results in the conductivity transformation. Besides the use of GO feature, this work also introduces conductive MXene to enhance fire detection speed and warning at low temperature, especially below 300 °C. The designed flame-retardant fire alarm is sensitive enough to detect fire incident, showing a response time of 2 s at 250 °C, which is calculated by a novel and quantifiable technique. More importantly, the designed fire alarm sensor is coupled to a wireless communication interface to conveniently transmit fire signal remotely. Therefore, when an abnormal temperature is detected, the signal is wirelessly transmitted to a liquid crystal display (LCD) screen when displays a message such as "FIRE DANGER". The designed smart fire alarm paper is promising for use as a smart wallpaper for interior house decoration and other applications requiring early fire detection and warning.

## 1. Introduction

Damage and injuries from fire accidents and hazards lead to significant loss of life and property globally each year. When fire disasters take place without control, they always cause irreparable massive casualties and huge property loss [1-3]. In order to mitigate or control fire, various researches focused on fire safety have been carried out. The current fire mitigation and fire protection methods include firefighting, the use of flame retardants and fire warning technologies [4-7]. Among these methods, the incorporation of flame retardants into material matrix is a common approach to improve flame retardancy. When flame-retardant materials encounter fire, the flame retardant additive acts to delay or extinguish the burning fire. Therefore, the flame retardant mainly functions at temperatures higher than 300 °C (i.e. during the combustion process). There are still some drawbacks caused by the use of additives in various materials, even though flame retardants play a positive role in the improvement of fire safety to a certain extent [8]. Consequently, the additives display poor compatibility with matrixes, leading to complicated preparation process and poor performance balance [9-11]. Besides, fire sensing methods have also been employed for timely fire detection and rapid response, which mainly works during pre-combustion process. Therefore, the early detection and warning of fire danger can increase the chances of rescue as well as time to put the fire out [12].

According to the different working mechanisms, early fire detection and warning systems are divided into smoke detectors and infrared detectors [8]. In general, smoke alarms can be triggered only after the continuous combustion of combustible materials [13], thus demonstrating that time is needed for smoke accumulation, which results in a relatively long response time, even more than 100 s [14-16]. Furthermore, most smoke alarms have low efficiency when they response in complicated outdoor environment, especially during rain and heavy wind situations [17]. Consequently, the fire-warning signal triggered by the aforementioned smoke alarms is not robust, fast and timely enough. Therefore, it is imperative to exploit more sensitive materials to design new fire-warning devices that can act reliably and timely providing real-time signals before the fire causes any damage. In response to the current challenges, several works have been focused on improving the sensitivity of fire alarms. Zhang. et al. prepared the GO based paper fire alarm by combining 3-methacryloxypropyltrimethoxysilane and L-ascorbic acid, which showed an improved flame retardancy and exhibited a rapid response time of 7 s at 300 °C [14]. In addition, polyurethane with decoration of ammonium polyphosphate and GO via electrostatic interactions was also used to construct a fire alarm, the alarm showed a response time of 11.2 s at 300 °C [18].

Moreover, a GO wide-ribbon wrapped sponges was also utilized to monitor fire, which exhibited a response time of 33 s to warn of the abnormal high temperature [8]. However, while the majority of the current research works employ GO, due to its temperature dependent electrical resistance transition effect [19, 20], the aforementioned works are limited by response temperatures above 300 °C and all the detected electric signals are converted into traditional warning such as light or a sound alarm signal to show fire danger. Therefore, as combustible materials usually get ignited, while exhibiting violent burning and rapid flame spread behavior [21-23] at temperatures rise above 300 °C, and to the best of our knowledge, there are only a few works on fire alarms focused on rapid response to fire at lower temperatures ( < 300 °C) with wireless signal transmission. Considering the convenience of wireless signal transmission, and urgency of detection of fire hazards at lower temperatures, it is necessary to design fire alarms with capabilities to rapidly detect fire at low temperature and transmit the electrical via a wireless communication system.

In this work, we developed a novel flame-retardant fire sensor decorated with GO and 2D material MXene with low temperature detection and wireless communication capabilities by employing bio-based cellulose paper as a substrate for the fire-warning system and phytic acid (PA) as flame retardant to keep structure intact in case of abnormal high temperature or fire. GO bears important role in fire-warning system, showing electrical insulation at room temperature and electrical conductivity when the temperature increases. MXene holding excellent water solubility ensures easy mixture with GO aqueous solution while also assisting PA to improve flame retardancy as well as accelerating the response time of the fire alarm system during fire scenarios at relatively low temperatures. Moreover, for the first time, a novel and precise method of fire alarm response detection is reported using a quantitative measure of light intensity. The designed fire alarm exhibits rapid fire detection response time at a low temperature of 178 °C and excellent remote warning and wireless signal transmission capability about 2 KM. This means fire alarm offers the longest fire-warning distance range. Our findings provide a promising method for the development of advanced materials for fire safety applications such as smart wallpapers.

## 2. Experimental

### 2.1 Raw materials

Phytic acid (PA) solution (50% (w/w) in water), hydrochloric acid (37%) (HCl) and lithium fluoride (LiF) were provided by Sigma-Aldrich, being used directly. Commercial $Ti_3AlC_2$ powder (MAX) was bought from Jilin 11 Technology Co., Ltd (Changchun, China). GO suspension was purchased from Grupo Antolin company. Deionized water was got from the water purification system in our laboratory. Cellulose paper was the common paper being cover on table.

### 2.2 Preparation of MXene

40 mL of HCl with a concentration of 9 M and 2 g LiF were put into the special polytetrafluoroethylene (PTFE) beaker and stirred for 30 min at 35 °C. Afterwards, 2 g MAX powder was added into above solution slowly. After reaction for 48 h, the obtained solution was centrifuged with deionized water, then ultrasonicated for 15 min. After PH value of upper liquid was more than 6, upper dark green suspension was collected. Lastly, freeze-dry the collected liquid to get exfoliated Ti$_3$C$_2$ powder.

## 2.3 Fabrication of fire-warning PA@GO and PA@MGOy papers

Firstly, cellulose paper was immersed into PA solution several times and dried at room temperature. Cellulose papers after PA modification were named as PA$_x$, where x was PA concentration. Next, the homogeneous MGO suspensions with different concentration ratio of GO to MXene were prepared for next usage. By using simple dip-coating way to put PAx samples into MGO suspension (GO suspension), final PA@MGOy (PA@GO) papers were obtained, where y meant concentration ratio of GO to MXene.

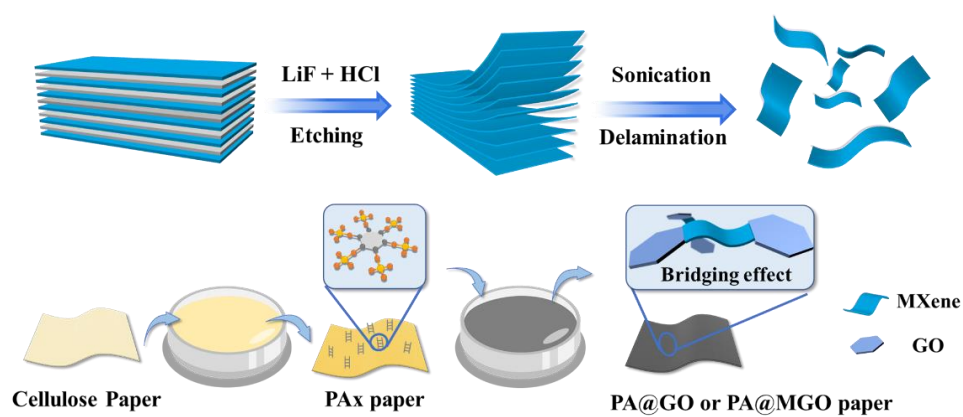

Scheme 1. Schematic fabrication process of PA@MGOy alarm samples.

Table 1. Components of PA@GO and PA@MGOy papers

| Samples | PA concentration/*wt%* | GO concentration/*(mg/mL)* | Concentration ration of GO to MXene |
| --- | --- | --- | --- |
| PA@GO | 20 | 2.5 | - |
| PA@MGO20 | 20 | 2.5 | 20:1 |
| PA@MGO15 | 20 | 2.5 | 15:1 |
| PA@MGO10 | 20 | 2.5 | 10:1 |
| PA@MGO5 | 20 | 2.5 | 5:1 |

## 2.1 Characterization

The combustion of the samples was investigated using the Microscale combustion calorimeter (MCC) (Fire Testing Technology, UK). The MCC is a small-scale instrument used to evaluate the combustion behavior by a standardized method (ASTM D7309-07). 5 mg of sample was heated form 100 °C to 700 °C with a heating rate of 1 °C/s to obtain the parameters of flame-retardancy. All samples were tested in triplicates. Vertical burning test with ignition time of 5 s was also used to evaluate the combustion behavior of the samples. Fourier transform infrared spectrometer (FTIR) was applied using Thermofisher Nicolet 5700 spectrometer in ATR mode. Thermogravimetric analysis (TGA) instrument (TA Q50) was also employed. 5-10 mg of samples were tested by heating from 50 °C to 750 °C at 10 °C/min under nitrogen atmosphere. Scanning electron microscopy (SEM) instrument (EVO MA15, Zeiss, Germany) was carried out to exam the structure. The conductive gold layer was sputtered on samples under 10 KV condition before observation. Transmission electron microscopy (TEM) (Tecnai T20, FEI) was also performed with EDS analysis in STEM mode. Electric resistance change was monitored by using multimeter (Keysight 3458A). X-ray diffraction (Philip X' Pert PRO diffractometer) with a Cu Kα radiation was used to obtain XRD pattern. Raman spectra was collected by Raman spectrometer (Renishaw PLC).

## 3. Results and discussion

### 3.1 Structure characterization of coated PA@GO and PA@MGOy paper

The successful and homogeneous coating is critical for samples' performance. For evaluating the coating, FTIR, SEM and EDS analyses were conducted and the corresponding results are shown in **Fig. 1**. From the FTIR spectra, it is observed that all the samples exhibit relatively sharp peaks at 3330 cm$^{-1}$ and 1025 cm$^{-1}$, corresponding to O-H stretching vibrations and C-O stretching vibrations, showing typical characteristic peaks in cellulose [24]. However, in contrast to pure cellulose paper, PA@GO and PA@MGOy samples exhibit weak characteristic peaks at 1630 cm$^{-1}$ and 1720 cm$^{-1}$ assigned to C=C from unoxidized sp$^2$ bonds and C=O stretching vibration of GO. This indicates the successful coating of GO on surface of cellulose paper due to hydrogen bond [25]. Additionally, it can also be seen from SEM figures that intact coating forms on surface of cellulose paper successfully. EDS test is also employed to further verify coating composition. Only carbon (C) and oxygen (O) elements are detected in PA@GO sample, while homogeneous titanium (Ti) element also appears in PA@MGO5 sample besides C and O elements, compared to PA@GO sample. The other PA@MGOy samples also show the similar EDS results with PA@MGO5 (**Figure S2**). This shows MXene also has been successful coated on surface of cellulose paper. Both of these results imply the successful coating on cellulose paper surface, which might be attributed to the hydrogen bonding [25].

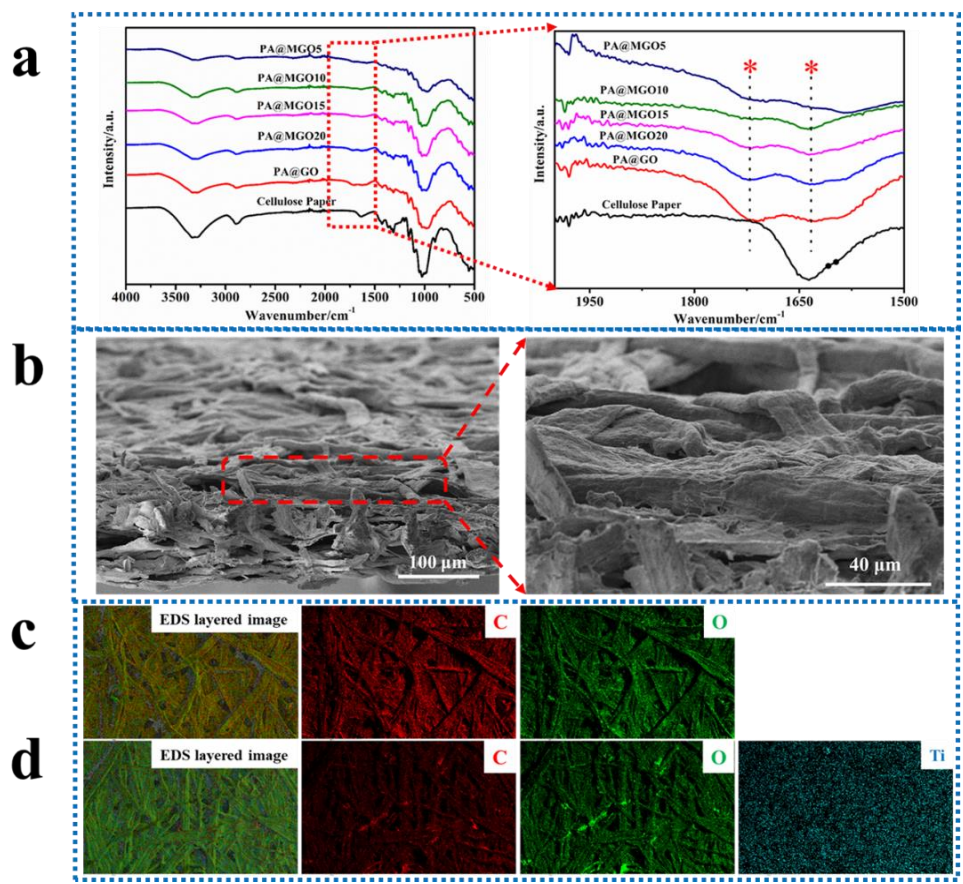

Figure 1. Structure characterization and analysis: (a) FTIR spectra of PA@GO and all PA@MGOy samples. (b) SEM observation of PA@MGO5. (c) EDS mapping of element in PA@GO. (d) EDS mapping of element in PA@MGO5.

**3.2 Flame-retardant property of fire-warning system**

Excellent thermal stability is crucial to maintain the samples structural integrity, which is the basis for fire-warning system. In order to evaluate the thermal stability of modified cellulose paper, TGA measurements were performed and the corresponding results are shown in **Fig. 2a**. While all the samples exhibit one stage decomposition, the residual weight of the different samples display clear difference, especially at high temperature. Accordingly, the residual weight of PA@MGOy increases with more load of MXene. PA@MGO5 sample shows relatively higher residual weight at 700 ºC, which indicates the improvement of flame retardancy for PA@MGOy as MXene content increases. Therefore, MXene addition can play a positive role in improving flame retardancy. It is likely due to the synergistic effect between MXene and PA, in which MXene can assist PA synergistically to inhibit heat release and isolate oxygen [26].

MCC test was also used to investigate the combustion behavior of the samples. The heat release profiles of all the samples show only one sharp peak, which implies a good homogenous coating. Moreover, the peak heat release rate (pHRR) gradually decreases after loading with MXene. In comparison to pure cellulose paper with pHRR value of 183 W/g, all PA@MGOy samples exhibit a lower pHRR value. Notably. the pHRR value of PA@MGO5 drops by 46.7 % to 97.5 W/g. These results demonstrate that MXene plays an essential role in improving flame retardancy. In this work, the combination of PA and MXene clearly results in better flame-retardant performance, which further proves the synergistic effect of MXene in the promotion of flame retardancy [26].

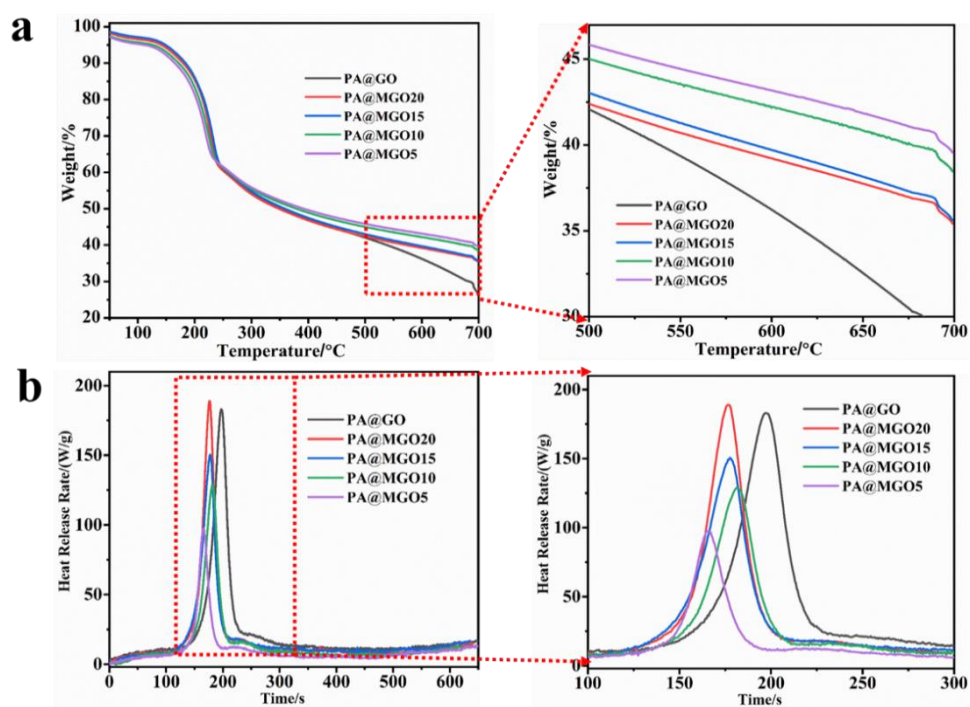

Figure 2. (a) TGA curves of PA@GO and PA@MGOy samples. (b) MCC curves of PA@GO and PA@MGOy samples.

### 3.3 Multi-response behavior of coated PA@GO and PA@MGOy paper

### 3.3.1 Electric resistance change vs heating time

The fire response of the fire-warning system mainly replies upon the transformation of GO from an insulator to a conductor during the heating process, which leads to samples' electrical resistance changes. So electric resistance changes as a function of time were measured at different temperatures (150 ºC, 200 ºC and 250 ºC), and the corresponding curves are provided in **Fig. 3a**. As expected, electric resistance of all samples sharply decreases when samples are exposed to heat. At the defined temperatures, the electrical resistance of the PA@GO sample takes relative longer time to decrease to lower value, which can support the transfer of electrons. Notably, after addition of MXene, electrical resistance changes of PA@MGO samples exhibit shorter time compared to that of PA@GO sample. In addition, the time for electrical resistance decrease exhibits further decrease with more loading of MXene content. At the same test temperature, the time for samples' electric resistance change to the value required to form an electrical circuit exhibit the following trend: PA@GO > PA@MGO20 > PA@MGO15 > PA@MGO10 > PA@MGO5. Interestingly, it is worth noting that electrical resistance change is strongly related to test temperature. Higher test temperature can lead to the faster electrical resistance change, which is due to the increased amount of heat energy at high test temperature, Thereby, leading to shorter time for electrical resistance transition to support fire-warning system. As test temperature increases from 150 ºC to 250 ºC, all the samples electrical resistance change time further shortens, which indicates increased sensitivity for the fire-warning system. Therefore, at the test temperature of 250 ºC, all the samples show the shortest time for electrical resistance change, hence displaying early fire detection.

**3.3.2 Fire-alarm response**

**3.3.2.1 Response time based on "Trigger Electrical Resistance"**

Normally, response time is set as one way to appraise the sensitivity for fire-warning system. One approach is to calculate response time of fire-warning system based on "trigger electrical resistance". Usually, when the electrical resistance decreases by over 90 %, the value is small enough to support the electrical circuit formation, that means fire alarm can be triggered immediately [12]. Hence, the time required to reach controlled trigger electrical resistance is selected to evaluate sensitivity of fire warning system. When electrical resistance reduces to trigger electrical resistance, the ratio of $R/R_0$ becomes 0.1. Therefore, the exact response time from original electrical resistance to trigger electrical resistance for fire alarm system is counted by this method, and the corresponding response time of each sample is displayed in **Fig. 3b**. At each test temperature, the PA@GO sample shows relatively the longest response time. After MXene introduction, the response time of PA@MGOy samples gradually decreases. With MXene content increases, the response time of PA@MGO further decreases. Moreover, at high test temperature, the response time decreases due to the increased heat energy. Compared to the response time at other test temperatures, the response time at 250 ºC is relatively shortest. And the response time at 250 ºC, follows the sequence: 24 s (PA@GO), 12 s (PA@MGO20), 10 s (PA@MGO15), 9 s (PA@MGO10) and 7 s (PA@MGO5), respectively. Comparatively, PA@MGO5 displays shortest response time of 7 s based "trigger electrical resistance" calculation method. Generally speaking, the trigger electrical resistance in this work is the lowest electrical resistance which can guarantee the electrical circuit formation. Therefore, it is possible for the electrical circuit to be formed when electrical resistance is slightly higher than the value of trigger electrical resistance selected in this work. To verify whether the above situation exists, another method is proposed in this work based on light intensity to calculate response time. The details of the method are shown in the section below.

**3.3.2.2 Response time based on "Light Intensity Change"**

In theory, once the transformation from insulation to conductive mode is realized, the electrical circuit will be formed and the lamp in circuit will be on. Therefore, when the light intensity is detected, it indicates fire-warning system has begun to work. This period of time until light intensity change is detected and set as response time, which shows the sensitivity of the fire-warning system. With reference to this calculation method, **Scheme 2** presents the detailed test method to obtain the quantified response time precisely. Fire-warning samples are put in an oven for heating. Once the coated GO on surface of sample is reduced to conductive RGO, the conductive circuit will turn the lamp on. When the lamp is on, the avaspec-2048L spectrometer machine will record the light intensity change. As expected, the collected light intensity increases with test time increase. When light intensity remains stable, which reaches to the maximum value. In this work, a sharp peak that appears at the wavelength of 529 nm when the lamp is on, signifies the response time. Therefore, the light intensity will change with test time, the corresponding curves of light intensity change at 529 nm as a function of time are shown in below **Fig. 3c**. And the response time calculated by "light intensity change" is also shown in **Fig. 3d**.

As seen in **Fig. 3c**, the light intensity change of each sample exhibits the similar trend when the electrical circuit is formed. It is clear that the light intensity gradually increases to a constant value over test time increase. Comparatively, the highest light intensity is exhibited by PA@MGO samples, which can be attributed to the good conductivity property of MXene. Therefore, the light intensity can be further enhanced with MXene content increase. It is a well-known fact that high temperature can accelerate reduction of GO coated on the surface of the samples, which explains why the light intensity change of every sample becomes faster as the test temperature increases (**Fig. 3c (i, ii and iii)**). When test temperature reaches to 250 ºC, all the samples take less time to form a conductive pathway.

In accordance with response time calculation method of "light intensity change", the time at which can capture the light intensity is key point for the fire-warning system. Therefore, this point at which light intensity changes is defined by making a tangent, and the corresponding time is set as response time for fire alarm. As shown in **Fig. 3d**, the change tendency of response time obtained by this method is consistent with the change trend of response time calculated by "trigger electric resistance". Similarly, the PA@GO sample exhibits a relatively longer response time by comparison with PA@MGOy samples. Once MXene is introduced on the surface of cellulose paper, the response becomes more rapid, showing shorter response time. As the test temperature increases from 150 ºC to 250 ºC, all the samples' response time gradually decreases. At the test temperature of 250 ºC, the response time of PA@MGOy samples is relatively shortest compared to the lower test temperatures. At this test temperature, PA@MGO5 sample exhibits the shortest response time of 2 s, showing that the coated cellulose paper is endowed with rapid electrical conductivity responsive behavior at a relatively lower test temperature.

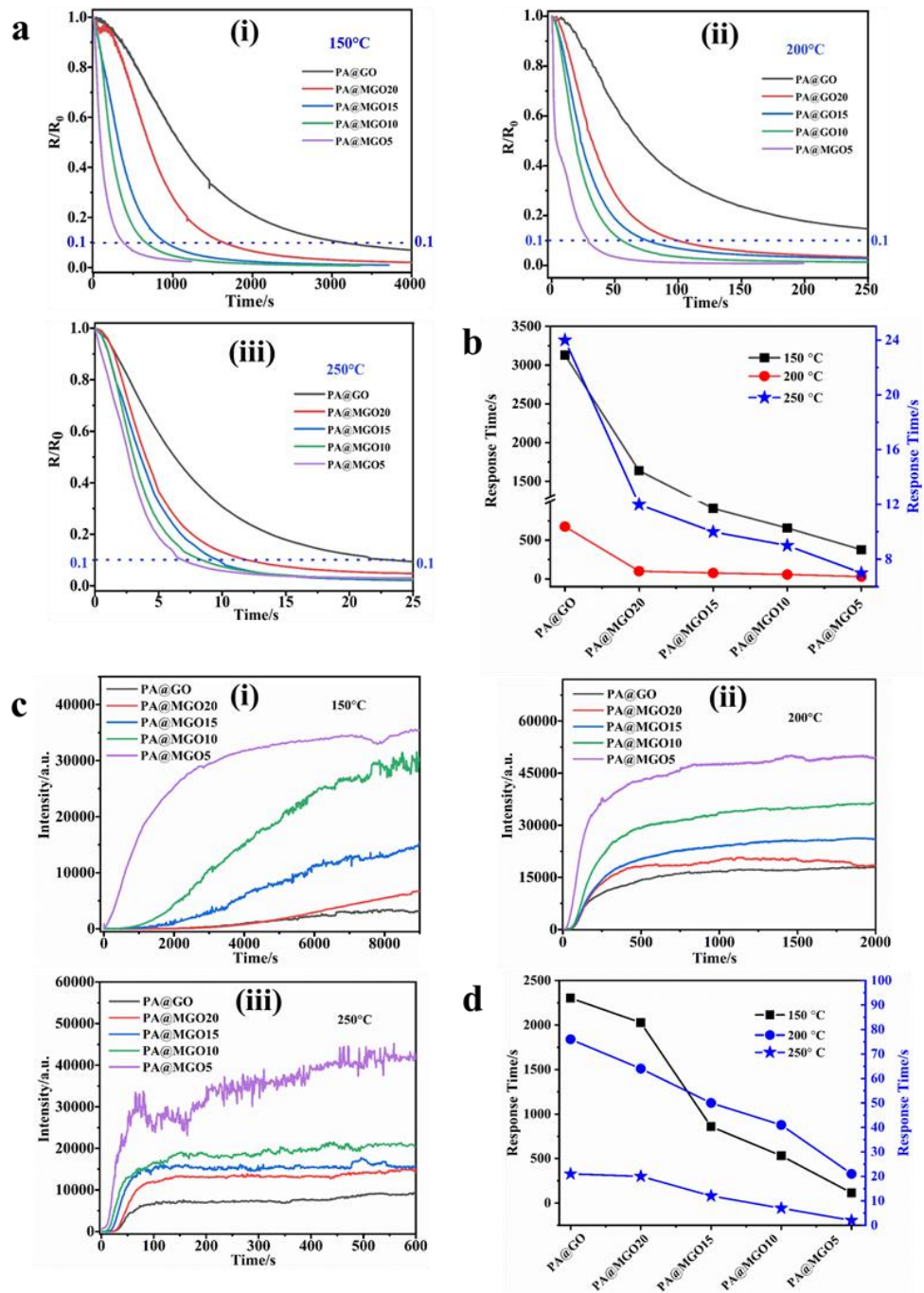

Figure 3. (a) Electrical resistance changes at different test temperature: (i) 150 ºC, (ii) 200 ºC and (iii) 250 ºC. (b) Response time calculated by "trigger electric resistance". (c) Curves of light intensity change versus different test time: (i) 150 ºC, (ii) 200 ºC and (iii) 250 ºC. (d) Response time calculated by "light intensity change".

As mentioned above, two methods are employed to calculate response time for fire-warning system. In order to make a better understanding for the distinction, a comparison of response time calculated by each method at different test temperature is exhibited in **Fig. 4a-4c**. At each test temperature, response time of different samples shows slight difference but the same tendency. Clearly, the response time obtained by "trigger electrical resistance" is relatively longer than one got by "lamp intensity change". The possible reason is the value of selected "trigger electrical resistance" is relatively lower than the one which can form circuit pathway. As previous explained, when the electrical resistance decreases to "trigger electrical resistance", which is about 90 % of electrical resistance decrease. Therefore, it is quite possible that at electrical resistance less than 90 %, a certain amount of RGO could be formed to facilitate electron transfer through the electrical circuit, which means the light intensity method can trigger the fire-warning system earlier. Comparatively, the obtained response time calculated by the light intensity change is more quantifiable and realistic. The response time of PA@MGO5 calculated by "light intensity change" is only 2 s at a test temperature of 250 ºC, which really shows advantages in low temperature rapid detection of fire-warning system. The response time obtained in this work is compared to the values reported in other fire alarm researches and summarized in **Fig. 4d** for better comparison. From the comparison, it can be seen that GO becomes the preferred materials for fire-warning systems because of its electrical resistance switching effect. Moreover, the fast response time mentioned in some researches occurs when test temperature is above 300 °C. There are a small number of studies involving fast response at low temperature, but the response time is long. Therefore, this work can quickly respond to fire signal at relative low temperatures, gaining more time for fire control.

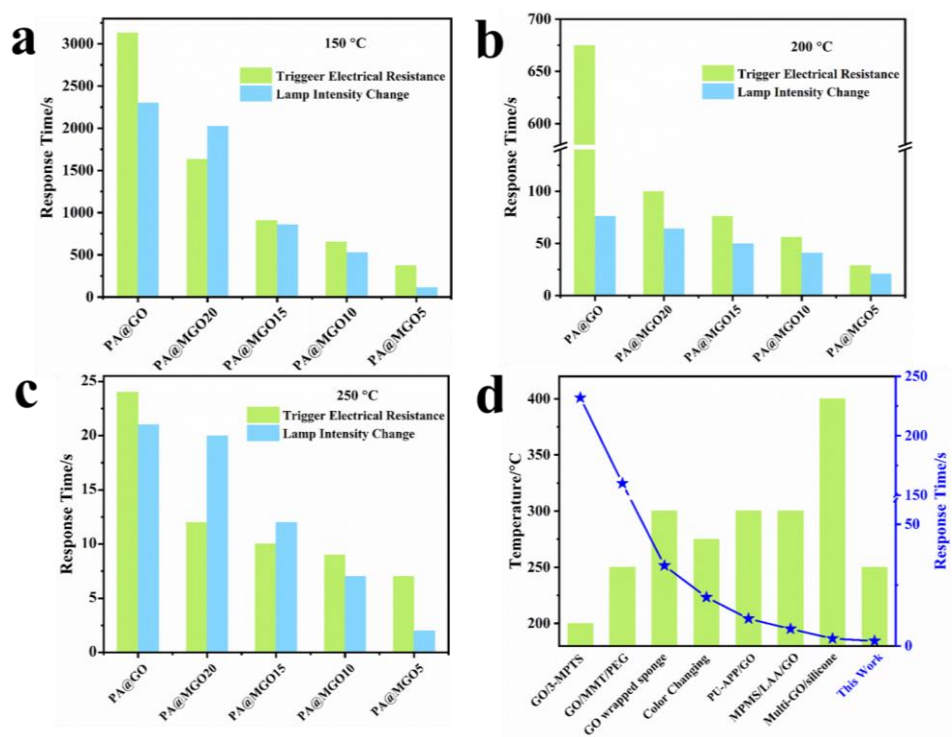

Figure 4. (a) Comparison of response time calculated by two methods at 150 ºC. (b) Comparison of response time calculated by two methods at 200 ºC. (c) Comparison of response time calculated by two methods at 250 ºC. (d) Comparison of alarm response time for other state-of-art fire-warning systems (Abbreviation: 3-mercaptopropyltrimethoxysilane (3-MPTS); montmorillonite (MMT); polyethylene glycol (PEG); ammonium polyphosphate (APP); polyurethane (PU); L-ascorbic acid (LAA); 3-methacryloxypropyltrimethoxysilane (MPMS); Multilayered graphene oxide (Multi-GO)).

### 3.4 Proposed temperature-responsive resistance transition mechanism

The key factor for fire-warning system is successful transition from electrical insulation at room temperature to electrical conductivity at high temperature. In this work, it is GO that undertake conductivity transformation. In order to verify the successful reduction of GO to RGO, proving the proposed working mechanism for fire-warning system. XRD and Raman tests are carried out and results are shown in **Fig. 5**. From XRD curve of GO, it is easy to see a sharp peak at 2θ ≈ 11 °, belonging to (001) reflection peak [27]. After GO reduction by heating process, (001) reflection peak disappears. There is a broad peak at 2θ ≈ 26 °, corresponding to the typical characteristic (002) graphene crystal plane of RGO [27]. In addition, Raman test is also used to detect structural change caused by heating reduction process. Both of the Raman spectrums include two strong characteristic peaks at ≈ 1348 cm$^{-1}$ (D band of carbon) and ≈ 1595 cm$^{-1}$ (G band of carbon) [28]. Relative intensity ratio $I_D/I_G$ values are different, indicating the change of graphitization degree. After heating treatment, $I_D/I_G$ value of PA@MGO sample increases from 0.59 to 0.78, which is due to an increase of surface defects content or edge area for RGO [27, 29-31]. It also indicates that most of oxygen-containing functional groups on GO surface are removed due to heating reduction process. Additionally, both D and G peaks in PA@MGO sample after heating become slightly sharp, showing decreased average size of sp$^2$ domain [30]. Based on above results, RGO has been successfully formed in the sensor samples after heating process, which provides conductivity for circuit.

On account of the analysis mentioned above, the proposed working mechanism for fire-warning system is revealed in **Fig. 5d**. The fire-warning alarm response is due to the fast thermal reduction process of GO nanosheets. Once fire or high temperature rise occurs, most of the oxygen-containing groups like carboxyl groups, epoxy groups and hydroxyl groups on surface of GO nanosheets experience thermal cracking and are removed from GO surface [32, 33]. This leads to the formation of RGO, which will form a conductive pathway on the GO surface. It is consistent with the above Raman and XRD results.

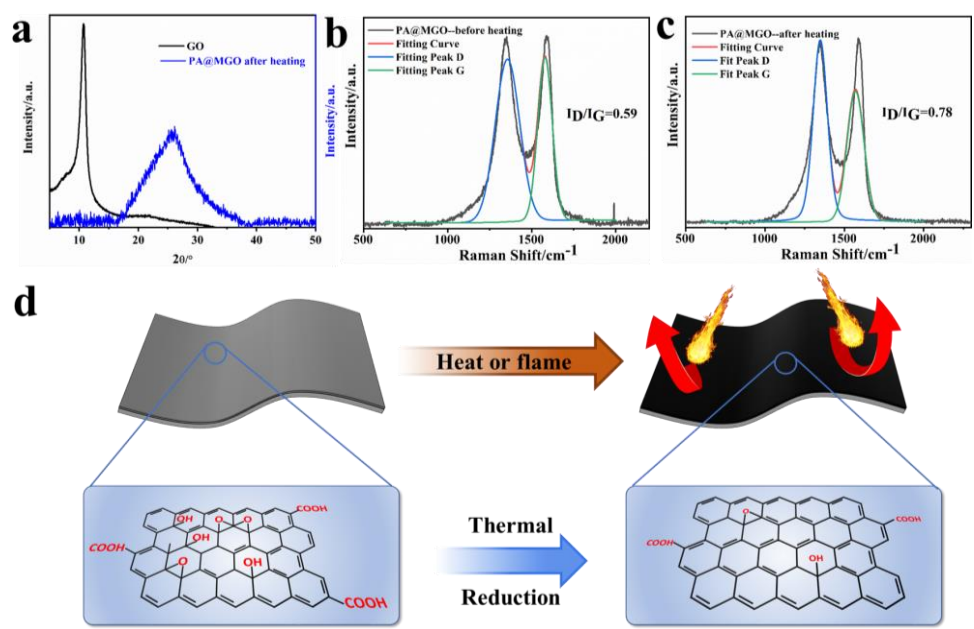

Figure 5. Proposed temperature-responsive resistance transition mechanism: (a) XRD curves of PA@MGO after heating. (b) Raman spectra of PA@MGO sample before heating. (c) Raman spectra of PA@MGO after heating. (d) Proposed working mechanism for fire-warning system.

**3.5 Wireless conversion from fire-warning signal**

**3.5.1 Wireless conversion device design**

With the aim of realizing remote monitoring fire situation to better control fire, a novel wireless signal conversion and communication device has been designed to process the signal obtained from the fire warning system. The design is mainly based on communication between a Wi-Fi transmitter (**Fig. 6a**) and a Wi-Fi receptor (**Fig. 6b**). PeakTech P6227 is the electrical power supply. It's two outputs are used as force electrodes: the positive (Force + ) is connected to one side of fire-warning paper and the negative one (Force -) is placed opposite to this electrode. Two sensing electrodes (Sense + and Sense -) are placed in the middle of paper, such that there is no contact between force electrodes. In addition, the ending parts of two sense electrodes are connected to the ground and analogue channel (i.e. A0) of the Arduino, which is programmed according to the application used. A threshold is established according to the material's response temperature, a fire warning message would be sent by Wi-Fi transmitter. The microcontroller of the Arduino board, used as analog-to-digital converter (ADC), reads the changing voltage and converts its continuous magnitude to a number between 0 and 1023, which is equivalent to 0 V and 5 V.

Two parameters of voltage supplied by source and threshold or number of steps (voltage measured) for ADC should be taken into account. If a maximum number 1023 is chosen, the temperature selected to activate the warning system would be the maximum measurable value for the voltage supply selected. Furthermore, because of the electrical noise existing in the board, the minimum number of steps corresponding to a minimum detection of voltage signal (maximum temperature) is 270. In our configuration, the voltage of the power supply used is 5 V and the number of steps is 870, corresponding to a voltage detected of $V = \frac{870 \times 5V}{1023} = 4,27\ V$ and the response temperature.

Maximum distance between Wi-Fi receptor and transmitter is approximately 2 km. Radio frequency signal is of 433 MHz. Several receptor-transmitter units could be positioned at different distances used as repeaters so that higher transmission distances could be obtained. Communication between Wi-Fi modules and channels is the following: DataTi-> DataRi, with DataT the transmitter channel, dataR, the data received by the receptor and i the number of channels going from 0 to 3. In addition, the receptor channel DataR0 is connected to an Arduino UNO with a Liquid crystal display (LCD) screen in this work, which showing "NORMAL" in normal state and "FIRE DANGER" (or any other message programmed) in abnormal temperatures. The LCD working principle is based on the Liquid Crystal Library which makes it easier to configure and control the characters displayed on the screen without dealing with low-level basics of how the screen actually works. Here, another threshold is programmed in the Arduino UNO board connected to the LCD: in case the analogue channel (i.e. A0) does not receive any signal, the message showed in the interface would be "NORMAL". However, if the threshold read by this channel is the maximum (1023, which corresponds to a voltage of 5V), the message displayed will change to "FIRE DANGER".

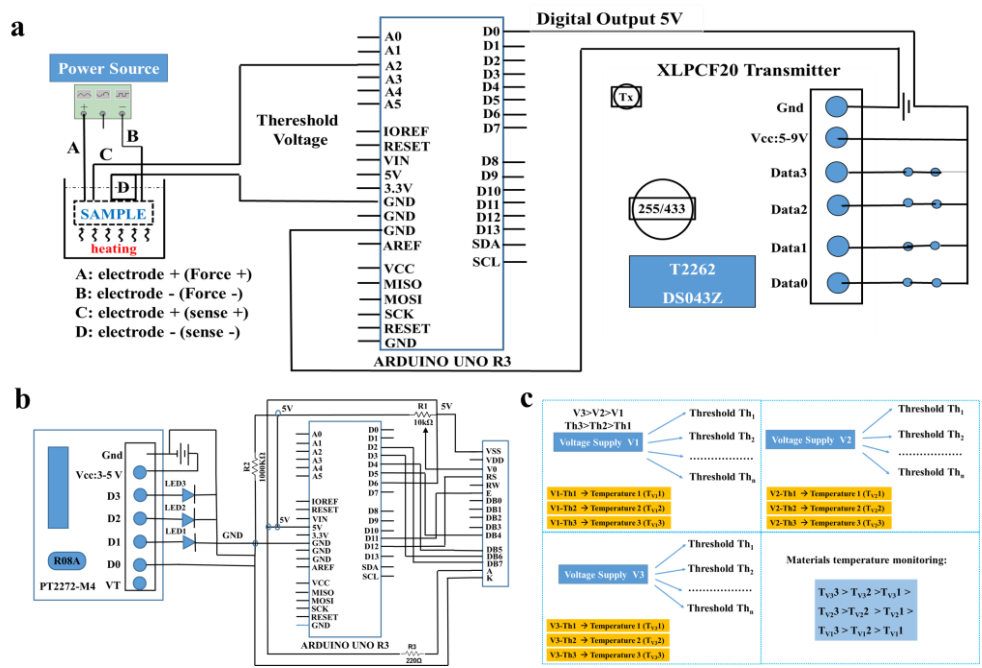

Figure 6. (a) Schematics of Wi-Fi ADC-transmitter for receiving alarm message and sending Wi-Fi signal. (b) Schematics of Wi-Fi receptor R080A and the Arduino connected to three LEDS and a LCD receiving above Wi-Fi signal for showing an alarm message in screen. (c) Working process description of the fire

warning system related to the threshold programmed in the ADC and voltage supply used.

### 3.5.2 Fire-alarm response process based on wireless conversion device

Schematic of fire warning system-to-wireless signal conversion and communication process for remote monitoring of fire process is presented in **Fig. 7**. Correspondingly, a video to show its function also is present in supporting information. Once the material is calibrated with the proper thresholds, the system may be placed at different locations in closed or open areas where continuous monitoring of environmental parameters is needed, such as the temperature supported by fire warning system presented in this work. As soon as the electrical circuit is formed after electrical resistance transformation of GO, due to a temperature increase or fire, a fire warning message will be sent to an operator who will be controlling the state of such parameters in another remote location. Moreover, another LCD display could be directly connected to the Wi-Fi emitter in the case a warning message is required at the location where fire or high temperature increase occurs. Besides, the remote fire warning radius can be extended using a LoRa internet of things (IoT) system and automated actuators such as fire extinguishers could be activated once the system warns of the danger.

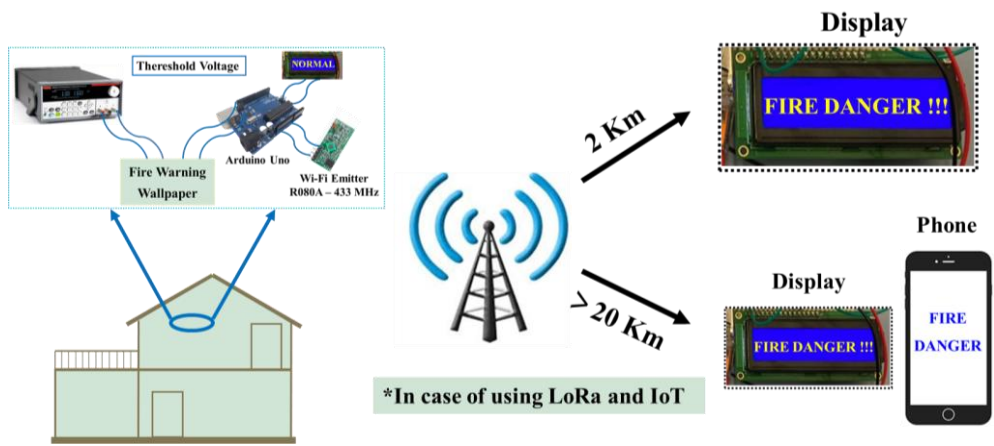

Figure 7. Schematic of wireless conversion from fire information to Wi-Fi signal for remote monitoring

fire alarm process.

## 4 Conclusion

A highly-sensitive fire-warning system based on PA@MGOy cellulose paper was successfully prepared via simple dip-coating method. After modification with two-dimensional GO and MXene, the flame-retardant PA@MGO5 paper exhibited a highly-sensitive response to abnormal temperature during pre-combustion process, showing a rapid response time of 2 s at a relatively low temperature of 250 °C, which will earn more time to ensure safe rescue and evacuation of people before fire development. A novel method to detect fire warning signal based on light intensity has been demonstrated to evaluate response time for the fire-warning system, which is more quantifiable and precise. It is noteworthy that a designed wireless signal conversion and communication device has been integrated to the fire-warning system for the first time, which successfully converts the received signal of the fire warning system into wireless signals and also displays a fire warning message on an LCD screen during high temperature or fire scenarios, within a radius of 2 km from incident location.

## Acknowledgements

This work is supported by China Scholarship Council, China under the Grant CSC (201908110272).

## Appendix A. Supplementary Data

Supporting information

## References

[1] B. Yuan, Y. Wang, G. Chen, F. Yang, H. Zhang, C. Cao, B. Zuo, Nacre-like graphene oxide paper

bonded with boric acid for fire early-warning sensor, J. Hazard. Mater. 403 (2021) 123645.

[2] D.M. Bowman, J.K. Balch, P. Artaxo, W.J. Bond, J.M. Carlson, M.A. Cochrane, C.M. D'Antonio, R.S. Defries, J.C. Doyle, S.P. Harrison, F.H. Johnston, J.E. Keeley, M.A. Krawchuk, C.A. Kull, J.B. Marston, M.A. Moritz, I.C. Prentice, C.I. Roos, A.C. Scott, T.W. Swetnam, G.R. van der Werf, S.J. Pyne, Fire in the earth system, Science 324(5926) (2009) 481-484.

[3] J.T. Randerson, H. Liu, M.G. Flanner, S.D. Chambers, Y. Jin, P.G. Hess, G. Pfister, M.C. Mack, K.K. Treseder, L.R. Welp, F.S. Chapin, J.W. Harden, M.L. Goulden, E. Lyons, J.C. Neff, E.A. Schuur, C.S. Zender, The impact of boreal forest fire on climate warming, Science 314(5802) (2006) 1130-1132.

[4] B. Yu, B. Tawiah, L.Q. Wang, A.C. Yin Yuen, Z.C. Zhang, L.L. Shen, B. Lin, B. Fei, W. Yang, A. Li, S.E. Zhu, E.Z. Hu, H.D. Lu, G.H. Yeoh, Interface decoration of exfoliated mxene ultra-thin nanosheets for fire and smoke suppressions of thermoplastic polyurethane elastomer, J. Hazard. Mater. 374 (2019) 110-119.

[5] C. Huang, X. Chen, B. Yuan, H. Zhang, H. Dai, S. He, Y. Zhang, Y. Niu, S. Shen, Suppression of wood dust explosion by ultrafine magnesium hydroxide, J. Hazard. Mater. 378 (2019) 120723.

[6] Y. Shi, C. Liu, L. Liu, L. Fu, B. Yu, Y. Lv, F. Yang, P. Song, Strengthening, toughing and thermally stable ultra-thin mxene nanosheets/polypropylene nanocomposites via nanoconfinement, Chem. Eng. J. 378 (2019) 122267.

[7] Y. Shi, C. Liu, L. Fu, F. Yang, Y. Lv, B. Yu, Hierarchical assembly of polystyrene/graphitic carbon nitride/reduced graphene oxide nanocomposites toward high fire safety, Compos. Part B: Eng. 179 (2019) 107541.

[8] H. Xu, Y. Li, N.J. Huang, Z.R. Yu, P.H. Wang, Z.H. Zhang, Q.Q. Xia, L.X. Gong, S.N. Li, L. Zhao, G.D. Zhang, L.C. Tang, Temperature-triggered sensitive resistance transition of graphene oxide wide-

ribbons wrapped sponge for fire ultrafast detecting and early warning, J. Hazard. Mater. 363 (2019) 286-294.

[9] G. Zu, K. Kanamori, A. Maeno, H. Kaji, K. Nakanishi, Superflexible multifunctional polyvinylpolydimethylsiloxane-based aerogels as efficient absorbents, thermal superinsulators, and strain sensors, Angew. Chem. Int. Ed. 57(31) (2018) 9722-9727.

[10] L. Dou, X. Zhang, X. Cheng, Z. Ma, X. Wang, Y. Si, J. Yu, B. Ding, Hierarchical cellular structured ceramic nanofibrous aerogels with temperature-invariant superelasticity for thermal insulation, ACS Appl. Mater. Interfaces 11(32) (2019) 29056-29064.

[11] G. Zu, K. Kanamori, T. Shimizu, Y. Zhu, A. Maeno, H. Kaji, K. Nakanishi, J. Shen, Versatile double-cross-linking approach to transparent, machinable, supercompressible, highly bendable aerogel thermal superinsulators, Chem. Mater. 30(8) (2018) 2759-2770.

[12] J. Chen, H. Xie, X. Lai, H. Li, J. Gao, X. Zeng, An ultrasensitive fire-warning chitosan/montmorillonite/carbon nanotube composite aerogel with high fire-resistance, Chem. Eng. J. 399 (2020) 125729.

[13] T. Fu, X. Zhao, L. Chen, W.-S. Wu, Q. Zhao, X.-L. Wang, D.-M. Guo, Y.-Z. Wang, Bioinspired color changing molecular sensor toward early fire detection based on transformation of phthalonitrile to phthalocyanine, Adv. Funct. Mater. 29(8) (2019) 1806586.

[14] Z.-H. Zhang, J.-W. Zhang, C.-F. Cao, K.-Y. Guo, L. Zhao, G.-D. Zhang, J.-F. Gao, L.-C. Tang, Temperature-responsive resistance sensitivity controlled by l-ascorbic acid and silane co-functionalization in flame-retardant go network for efficient fire early-warning response, Chem. Eng. J. 386 (2020) 123894.

[15] D.D. EVANS, a.D.W. STROUP, Methods to calculate te response time of heat and smoke detectors

installed below large unobstructed ceilings, Fire Technol. 22 (1986) 54-65.

[16] J.R.Q. III, Fire test comparison of smoke detector response times, Fire Technol. 36 (2000) 89-102.

[17] K.J. Overholt, M.J. Gollner, J. Perricone, A.S. Rangwala, F.A. Williams, Warehouse commodity classification from fundamental principles. Part ii: Flame heights and flame spread, Fire Safety J. 46(6) (2011) 317-329.

[18] K.-Y. Guo, Q. Wu, M. Mao, H. Chen, G.-D. Zhang, L. Zhao, J.-F. Gao, P. Song, L.-C. Tang, Water-based hybrid coatings toward mechanically flexible, super-hydrophobic and flame-retardant polyurethane foam nanocomposites with high-efficiency and reliable fire alarm response, Compos. Part B: Eng. 193 (2020) 108017.

[19] D. Li, M.B. Muller, S. Gilje, R.B. Kaner, G.G. Wallace, Processable aqueous dispersions of graphene nanosheets, Nat. Nanotechnol. 3(2) (2008) 101-105.

[20] S.K. Kim, J.Y. Kim, B.C. Jang, M.S. Cho, S.-Y. Choi, J.Y. Lee, H.Y. Jeong, Conductive graphitic channel in graphene oxide-based memristive devices, Adv. Funct. Mater. 26(41) (2016) 7406-7414.

[21] N.-J. Huang, Q.-Q. Xia, Z.-H. Zhang, L. Zhao, G.-D. Zhang, J.-F. Gao, L.-C. Tang, Simultaneous improvements in fire resistance and alarm response of go paper via one-step 3-mercaptopropyltrimethoxysilane functionalization for efficient fire safety and prevention, Compos. Part A: Appl. S. 131 (2020) 105797.

[22] S.T. McKenna, N. Jones, G. Peck, K. Dickens, W. Pawelec, S. Oradei, S. Harris, A.A. Stec, T.R. Hull, Fire behaviour of modern facade materials - understanding the grenfell tower fire, J. Hazard. Mater. 368 (2019) 115-123.

[23] Z.-R. Yu, M. Mao, S.-N. Li, Q.-Q. Xia, C.-F. Cao, L. Zhao, G.-D. Zhang, Z.-J. Zheng, J.-F. Gao, L.-C. Tang, Facile and green synthesis of mechanically flexible and flame-retardant clay/graphene oxide

nanoribbon interconnected networks for fire safety and prevention, Chem. Eng. J. 405 (2021) 126620.

[24] V. Hospodarova, E. Singovszka, N. Stevulova, Characterization of cellulosic fibers by ftir spectroscopy for their further implementation to building materials, Am. J. Anal. Chem. 09(06) (2018) 303-310.

[25] G. Chen, B. Yuan, Y. Wang, X. Chen, C. Huang, S. Shang, H. Tao, J. Liu, W. Sun, P. Yang, G. Shi, Nacre-biomimetic graphene oxide paper intercalated by phytic acid and its ultrafast fire-alarm application, J. Colloid Interface Sci. 578 (2020) 412-421.

[26] Y. Xue, J. Feng, S. Huo, P. Song, B. Yu, L. Liu, H. Wang, Polyphosphoramide-intercalated mxene for simultaneously enhancing thermal stability, flame retardancy and mechanical properties of polylactide, Chem. Eng. J. 397 (2020).

[27] M. Kim, C. Lee, J. Jang, Fabrication of highly flexible, scalable, and high-performance supercapacitors using polyaniline/reduced graphene oxide film with enhanced electrical conductivity and crystallinity, Adv. Funct. Mater. 24(17) (2014) 2489-2499.

[28] F. Niu, J. Yang, N. Wang, D. Zhang, W. Fan, J. Yang, Y. Qian, Mose2-covered n,p-doped carbon nanosheets as a long-life and high-rate anode material for sodium-ion batteries, Adv. Funct. Mater. 27(23) (2017) 1700522.

[29] R. Furlan de Oliveira, P.A. Livio, V. Montes-García, S. Ippolito, M. Eredia, P. Fanjul-Bolado, M.B. González García, S. Casalini, P. Samorì, Liquid-gated transistors based on reduced graphene oxide for flexible and wearable electronics, Adv. Funct. Mater. 29(46) (2019) 1905375.

[30] A.C. Ferrari, D.M. Basko, Raman spectroscopy as a versatile tool for studying the properties of graphene, Nat. Nanotechnol. 8(4) (2013) 235-246.

[31] P. Yao, P. Chen, L. Jiang, H. Zhao, H. Zhu, D. Zhou, W. Hu, B.H. Han, M. Liu, Electric current

induced reduction of graphene oxide and its application as gap electrodes in organic photoswitching devices, Adv. Mater. 22(44) (2010) 5008-5012.

[32] H. Xie, X. Lai, H. Li, J. Gao, X. Zeng, X. Huang, S. Zhang, A sandwich-like flame retardant nanocoating for supersensitive fire-warning, Chem. Eng. J. 382 (2020) 122929.

[33] F.F. Chen, Y.J. Zhu, F. Chen, L.Y. Dong, R.L. Yang, Z.C. Xiong, Fire alarm wallpaper based on fire-resistant hydroxyapatite nanowire inorganic paper and graphene oxide thermosensitive sensor, ACS Nano 12(4) (2018) 3159-3171.